\documentclass [preprint,showpacs,nofootinbib,superscriptaddress]{revtex4}
\usepackage{verbatim}
\usepackage{slashed}
\usepackage{epsfig}
\usepackage{feynmf}
\usepackage{graphics}
\usepackage{amsmath}
\usepackage{mathrsfs}
\usepackage{bm}

\begin{document}


\title{Search for $C=+$ charmonium and bottomonium states in
$e^+e^-\to \gamma+~X$ at B factories}


\author{Dan Li\footnote{Present address: Department of Physics, University
of Wisconsin, Madison, WI53706, USA}}
\affiliation{Department of
Physics and State Key Laboratory of Nuclear Physics and Technology,
Peking University, Beijing 100871, China}
\author{Zhi-Guo He}
\affiliation{Department of Physics and State Key Laboratory of
Nuclear Physics and Technology, Peking University, Beijing 100871,
China}

\affiliation{\small{\it{Departament d'Estructura i Constituents de
la Mat\`eria
                   and Institut de Ci\`encies del Cosmos}}\\
        \small{\it{Universitat de Barcelona}}\\
        \small{\it{Diagonal, 647, E-08028 Barcelona, Catalonia, Spain.}} }

\author{Kuang-Ta Chao}
\affiliation{Department of Physics and State Key Laboratory of
Nuclear Physics and Technology, Peking University, Beijing 100871,
China}


\begin{abstract}
We study the production of $C=+$ charmonium states $X$ in $e^+e^-\to
\gamma~+~X$ at B factories with $X=\eta_c(nS)$ (n=1,2,3),
$\chi_{cJ}(mP)$ (m=1,2), and $^1D_2(1D)$. In the S and P wave case,
contributions of QED with one-loop QCD corrections are calculated
within the framework of nonrelativistic QCD(NRQCD) and in the D-wave
case only the QED contribution is considered. We find that in most
cases the one-loop QCD corrections are negative and moderate, in
contrast to the case of double charmonium production $e^+e^-\to
J/\psi~+~X$, where one-loop QCD corrections are positive and large
in most cases. We also find that the production cross sections of
some of these states in $e^+e^-\to \gamma~+~X$ are larger than that
in $e^+e^-\to J/\psi~+~X$ by an order of magnitude even after the
negative one-loop QCD corrections are included. We then argue that
search for the X(3872), X(3940), Y(3940), and X(4160) in $e^+e^-\to
\gamma~+~X$ at B factories may be helpful to clarify the nature of
these states. For completeness, the production of bottomonium states
in $e^+e^-$ annihilation is also discussed.
\end{abstract}


\maketitle

\section{Introduction}
In recent years there have been a number of exciting discoveries of
new hidden charm states, i.e. the so called XYZ mesons, by Belle,
BaBar, CLEO, CDF, and D0 collaborations (for recent experimental and
theoretical reviews and related references see Ref.\cite{rev}).
Among the XYZ states, the charge parity C=+ states such as X(3872),
X(3940), Y(3940), Z(3930), and X(4160) are particularly interesting
and the interpretations for their nature are still very inconclusive
(except for the Z(3930), which is assigned as the $\chi_{c2}(2P)$
meson). The experimental results for these C=+ states have induced
renewed theoretical interest in understanding the mass spectrum,
decay and production mechanisms of charmonium or charmoniumlike
states (see, e.g., Refs.\cite{rev,10,11,12,Li09}). Among others, the
double charmonium production in $e^+e^-$ annihilation at
B-factories\cite{Abe:2002rb,BaBar:2005} turned out to be a good way
to find the C=+ charmonium or charmoniumlike states, recoiling
against the easily reconstructed $1^{--}$ charmonium $J/\psi$ and
$\psi(2S)$. In addition to the $\eta_c, \eta_c(2S)$ and $\chi_{c0}$,
the $X(3940)$ (decaying into $D\bar {D^*}$) and $X(4160)$ (decaying
into $D^*\bar {D^*}$) have also been observed in double charmonium
production. Since the quantum number of the photon is the same as
$J/\psi$, it will be interesting to see whether the C=+ charmonium
or charmoniumlike states can be found in the process $e^+e^-\to
\gamma ^*\to \gamma+X$, where $X$ is a C=+ state recoiling against
the photon. The production rates of such processes have been
calculated at tree level in QED\cite{Lee08}.

It has been known for some time that the one-loop QCD radiative
corrections are very important in double charmonium production in
$e^+e^-$ annihilation. The observed double charmonium production
cross section\cite{Abe:2002rb,BaBar:2005} for $e^+e^-\to
J/\psi\eta_c$ is larger than the leading-order (LO) calculations in
NRQCD\cite{Bodwin:1994jh} by an order of
magnitude\cite{Braaten:2002fi}, and later it was found that these
discrepancies could be largely resolved by the next-to-leading-order
(NLO) QCD corrections\cite{Zhang:2005cha,Gong:2007db} combined with
relativistic corrections\cite{bodwin06,He:2007te}. Therefore, it is
necessary to examine whether the one-loop QCD (i.e., $O(\alpha_s)$)
corrections are also important for the processes $e^+e^-\to \gamma
^*\to \gamma+X$. In fact, the one-loop QCD radiative correction to
$e^+e^-\to \gamma ^*\to \gamma+\eta_c$ has been investigated
elsewhere\cite{Shifman:1980dk,jia}.

Another interesting point is about $1^{++}$ charmonium. At $B$
factories the observed production cross sections in $e^+e^-$
annihilation to $J/\psi\eta_c$, $\psi(2S)\eta_c$,
$J/\psi\eta_c(2S)$, $J/\psi\chi_{c0}$, and $\psi(2S)\chi_{c0}$ are
large, but no signals for $J/\psi\chi_{c1,2}$ have been seen.
This is in line with the calculations in NRQCD\cite{Braaten:2002fi},
in which the predicted production rates of $J/\psi\chi_{c1,2}$ are
relatively suppressed. We wonder whether the cross section of
$1^{++}$ charmonium (including $\chi_{c1}$ and its radial
excitations) associated with a photon could be large in $e^+e^-\to
\gamma ^*\to \gamma+X$. If this is the case, we might have a chance
to search for the $\chi_{c1}$ as well as the X(3872) in $e^+e^-\to
\gamma ^*\to \gamma+X$, since the X(3872) could be a $\chi_{c1}(2P)$
dominated state but mixed with some $D^0\bar {D^{*0}}$ component, in
one of the possible interpretations. This is also useful to the
search for the Y(3940), which has been seen in the decay $B\to
Y(3940)K$ followed by $Y(3940)\to J/\psi\omega$, and is also a
possible candidate for the $\chi_{c1}(2P)$ (or $\chi_{c0}(2P)$). Of
course, these states could have some more exotic nature, being
molecules, tetraquarks, or charmonium hybrids.

In this paper, we compute the QED (at tree level) and one-loop QCD
($O(\alpha_s)$) corrections to the processes $e^+e^-\to \gamma
^*\to\gamma+X$, where X are $\chi_{cJ},\eta_c,{}^1D_2$ and their
radially excited states, all with charge-parity $C=+1$.  We find the
cross sections for $\eta_c$, its radial excited states and
$\chi_{c1}$, $\chi_{c1}(2P)$ are relatively large. Despite of the
large background from initial state radiation (ISR), we still expect
they could be seen in the $\gamma$ recoil spectrum with higher
statistics in the future. The remainder of the paper is organized as
follows. In Sec.2  we outline the QED calculation and some basic
techniques for numerically computing the one-loop QCD correction.
The QED and one-loop QCD corrections to cross sections for
$e^+e^-\to \gamma+X$ at B factories are given in Sec. 4, and we also
analyze and discuss our results. In the Appendix, we show some basic
integration expressions.

\section{QED Calculation}

\begin{figure}[htb]
\begin{center}
\scalebox{1}{\includegraphics{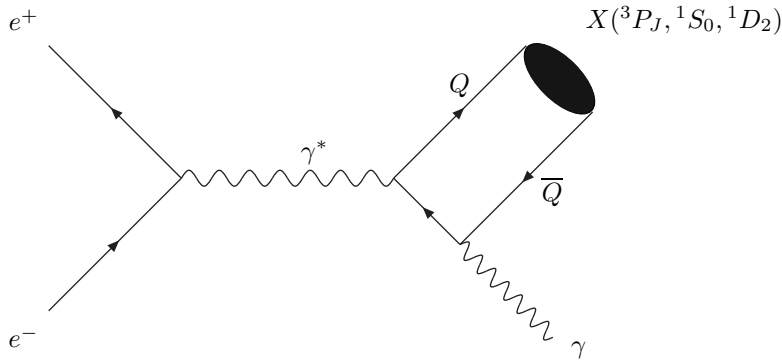}}
\end{center}
\caption{The tree QED diagram for $e^+e^-\longrightarrow \gamma+X$}
\end{figure}

The Feynman diagram for the exclusive process $e^+e^-\longrightarrow
\gamma+X$ at order $\alpha^3\alpha_s^0$ is shown in Fig.1, where X
is a heavy quarkonium with charge-parity $C=+1$, and there is
another quark line-flipped one. In the nonrelativistic limit, the
factorization formula for heavy quarkonium production in the NRQCD
framework is equivalent to that in the color-singlet model. And in
our case, the amplitude $\mathcal{M}$ for $e^{+}e^{-}\to\gamma+X$
can be expressed as
\begin{eqnarray}
&&\mathcal{M}(e^{+}e^{-}\to\gamma+X)=\sum_{S,L}\sum_{s_1,s_2}\sum_{i,j}
\int
\frac{{d}^3\mathbf{q}}{(2\pi)^{3}2q^{0}}\delta(q^{0}-\frac{\mathbf{q}^2}{2m_Q})\psi(\mathbf{q})\langle
s_1,\hspace{-0.1cm}s_2|SS_z\rangle\nonumber\\
&&\langle LL_z,SS_z|JJ_z\rangle\langle i,j|1\rangle
\mathcal{A}(e^{+}e^{-}\rightarrow\gamma+
Q_{s_1}^{i}(\frac{P}{2}+q)+\overline{Q}_{s_2}^{j}(\frac{P}{2}-q))
\end{eqnarray}
where $P$ is the momentum of $X$ state, $2q$ is relative momentum
between $Q$ and $\overline{Q}$ in the rest frame of X state, and
$\langle LL_z;SS_z|JJ_z\rangle$, $\langle s_1;s_2|SS_z\rangle$ and
$\langle i,j|1\rangle=\delta_{i,j}/\sqrt{N_c}$ are the spin-SU(2),
angular momentum C-G coefficients and color-SU(3) C-G coefficients
for $Q\bar{Q}$ pairs projecting onto appropriate bound states
respectively. And $\mathcal{A}$ is the standard Feynman amplitude
denoting $e^{+}e^{-}\to\gamma+
Q_{s_1}^{i}(\frac{P}{2}+q)+\overline{Q}_{s_2}^{j}(\frac{P}{2}-q)$.

The Feynman amplitude part can be evaluated by introducing the spin
projection operator\cite{Kuhn:1979bb,20}:
\begin{equation}
P_{SS_z}(P,q)\equiv\sum\limits_{s_1s_2 }\langle s_1;s_2|SS_z\rangle
v(\frac{P}{2}-q;s_2)\bar{u}(\frac{P}{2}+q;s_1).
\end{equation}
Expanding the operator in terms of the relative momentum $q$, 
we get the leading-order nonvanishing terms for the $S$-, $P$- and $D$-wave
case respectively. The results of the spin-triplet and spin-singlet
projection operators and their derivatives with respective to the
relative momentum $q_{\alpha}$ are given below\cite{Ko:1996xw}:
\begin{eqnarray} \label{pjs} P_{1S_z}(P,0)&=&\frac{1}{2\sqrt{2}}\
\slashed{\epsilon}^{\ast}(S_z)(\slashed{P}+2m_Q), \\
\label{der} P_{1S_z}^{\alpha}(P,0)&=&\frac{1}{4\sqrt{2}m_Q}
[\gamma^{\alpha}\slashed{\epsilon}^*(S_z)(\slashed{P}+2m_Q)-
(\slashed{P}-2m_Q)\slashed{\epsilon}^{\ast}(S_z)\gamma^{\alpha}]. \\
\label{petc}
P_{00}(P,0)&=&\frac{1}{2\sqrt{2}}\gamma_5(\slashed{P}+2m_Q),
\\P_{00}^{\alpha\beta}(P,0)&=&\frac{1}{8\sqrt{2}m_Q^2}(\gamma^\alpha
(\slashed{P}-2m_Q)\gamma^\beta+\gamma^\beta
(\slashed{P}-2m_Q)\gamma^\alpha)\gamma^5+vanishing \ \ terms
\end{eqnarray}
After integrating $q^0$, we get the amplitudes for $S$, $P$ and
$D$-wave heavy quarkonium production respectively:
\begin{equation}
\mathcal{M}(\gamma+\eta_c)=\int\frac{{d}^3\mathbf{q}}{(2\pi)^3}\psi_{00}(\textbf{q})\mathrm{Tr}[P_{00}O]|_{q=0}
\end{equation}
\begin{equation}
\mathcal{M}(\gamma+\chi_{cj})=\int\frac{{d}^3\mathbf{q}}{(2\pi)^3}q_\alpha\psi_{1m}(\mathbf{q})\langle
LL_z;SS_z|JJ_z\rangle\varepsilon^{\ast}_\beta(S_z)\mathrm{Tr}[P_{1S_z}^\beta
O^\alpha+P_{1S_z}^{\beta\alpha}O]|_{q=0}
\end{equation}
\begin{equation}
\mathcal{M}(\gamma+{}^1D_2)=\int\frac{{d}^3\mathbf{q}}{(2\pi)^3}\frac{1}{2}q_\alpha
q_\beta\psi_{2m}(\mathbf{q})
\mathrm{Tr}[P_{00}^{\alpha\beta}O+P_{00}^\alpha O^\beta+P_{00}^\beta
O^\alpha+P_{00}O^{\alpha\beta}]|_{q=0}
\end{equation}
where $O$ is the $\gamma$ matrix relevant to the Feynman amplitude
$\mathcal{A}$, and $O^{\alpha}$ and $O^{\alpha\beta}$ are the first and
second derivatives of $O$ with respect to $q_{\alpha}$ respectively.

The integrals of the wave function in momentum space are related to
the radial wave function $R_S(0)$, $R_P^{\prime}(0)$ and
$R_D^{\prime\prime}(0)$ in coordinator space at the origin for the $S-$,
$P-$ and $D-$ wave cases respectively:
\begin{equation}
\int\frac{d^3 \mathbf{q}}{(2\pi)^3}\psi_{00}(\mathbf
q)=\frac{1}{\sqrt{4\pi}}R_S(0),
\end{equation}
\begin{equation}
 \int\frac{d^3 \mathbf{q}}{(2\pi)^3}q_{\alpha}
 \psi_{1m}({\bf q})=-i\epsilon^{\ast}_{\alpha}(L_z)\sqrt{\frac{3}{4\pi}}R_P^{\prime}(0),
\end{equation}
\begin{equation}
 \int\frac{d^3 q}{(2\pi)^3}q_{\alpha}q_{\beta}
 \psi_{2m}({\bf q})=\varepsilon^{\ast}_{\alpha\beta}(L_z)\sqrt{\frac{15}{8\pi}}
 R_D^{\prime\prime}(0),
\end{equation}
where $\epsilon^{\alpha}(L_z)$ is the polarization vector of $L=1$
($P$-wave) system and $\varepsilon_{\alpha\beta}^m(L_z)$ is the
polarization tensor of $L=2$ ($D$-wave) system.

For spin-triplet $P$-wave states, the projection of the $L-S$
coupling of the spin vector $\epsilon^{\ast}(S_z)$ and orbital
vector $\epsilon^{\ast}(L_z)$ onto total angular momentum $J$ for
$J=0,1,2$ are
\begin{subequations}
\begin{equation}
\epsilon^{\ast}_\alpha(S_z)\epsilon^{\ast}_{\beta}(L_z)\langle
1,L_z;\;1S_z|0,0\rangle=\frac{1}{\sqrt{3}}\Pi_{\alpha\beta},
\end{equation}
\begin{equation}
\epsilon^{\ast}_\alpha(S_z)\epsilon^{\ast}_{\beta}(L_z)\langle
1,L_z;\;1S_z|1,J_z\rangle=\frac{i}{2\sqrt{2}m_Q}\epsilon^{\alpha\beta\rho\kappa}
P_{\kappa}\epsilon^{\ast}(J_z),
\end{equation}
\begin{equation}
\epsilon^{\ast}_\alpha(S_z)\epsilon^{\ast}_{\beta}(L_z)\langle
1,L_z;\;1S_z|2,J_z\rangle=\varepsilon^{\ast}_{\alpha\beta},
\end{equation}
\end{subequations}
where $\Pi_{\alpha\beta}=(-g_{\alpha\beta}+
\frac{P_{\alpha}P_{\beta}}{4m_Q^2})$. For the total angular momentum
$J=1$ and $J=2$ states, the sums over all possible polarizations are
given by
\begin{subequations}
\begin{align}
\sum_{J_z}\epsilon_{\alpha}(J_z)\epsilon^{\ast}_{\beta}(J_z)=\Pi_{\alpha\beta},
\end{align}
\begin{align}
\sum_{J_z}\varepsilon_{\alpha\beta}(J_z)\varepsilon^{\ast}_{\alpha^{\prime}\beta^{\prime}}(J_z)
=1/2(\Pi_{\alpha\alpha^{\prime}}\Pi_{\beta\beta^{\prime}}+
\Pi_{\alpha\beta^{\prime}}\Pi_{\alpha^{\prime}\beta})-
\frac{1}{3}\Pi_{\alpha\beta}\Pi_{\alpha^{\prime}\beta^{\prime}}.
\end{align}
\end{subequations}
With the help of the formula introduced above, we get the final QED
analytic expressions for the exclusive process
$e^+e^-\longrightarrow \gamma+X$:
\begin{subequations}
\begin{eqnarray}
\sigma(e^+e^-\rightarrow\gamma+\eta_c)=\frac{3\alpha^3e_c^4|R_S(0)|^2(1-r)}{s^2m_c}\int
d\Omega(1+\cos^2(\theta))
\end{eqnarray}
\begin{eqnarray}
\sigma(e^+e^-\rightarrow\gamma+\chi_{c0})=\frac{3\alpha^3e_c^4|R^{\prime}_P(0)|^2(1-3r)^2}{s^2m_c^3(1-r)}\int
d\Omega(1+\cos^2(\theta))
\end{eqnarray}
\begin{eqnarray}
\sigma(e^+e^-\rightarrow\gamma+\chi_{c1})=\frac{18\alpha^3e_c^4|R^{\prime}_P(0)|^2}{s^2m_c^3(1-r)}\int
d\Omega(1+2r+(1-2r)\cos^2(\theta))
\end{eqnarray}
\begin{eqnarray}
\sigma(e^+e^-\rightarrow\gamma+\chi_{c2})=\frac{6\alpha^3e_c^4|R^{\prime}_P(0)|^2}{s^2m_c^3(1-r)}\int
d\Omega(1+6r+6r^2+(1-6r+6r^2)\cos^2(\theta))
\end{eqnarray}
\begin{eqnarray}
\sigma(e^+e^-\rightarrow\gamma+{}^1D_2)=\frac{15\alpha^3e_c^4|R^{\prime\prime}_D(0)|^2(1-r)}{s^2m_c^5}\int
d\Omega(1+\cos^2(\theta))
\end{eqnarray}
\end{subequations}
where $r=M_x^2/s$, $\cos(\theta)$ is the angle between $J/\psi$ and
the initial beam axis. For the $c\bar{c}$ system we set $M_X=2m_c$. If
we replace $\frac{3}{4\pi}|R^{\prime}_p(0)|^2$ by
$\frac{1}{N_c^2-1}\frac{\langle O_8({}^3P_J)\rangle}{2J+1}$, we find
our QED results of ${}^3P_J$ are consistent with those in
Ref.\cite{Braaten:1995ez}. For $b\bar{b}$ states, the result can be
obtained by changing $e_c$ to $e_b$, $m_c$ to $m_b$ and the values
of the wave-functions for charmonium states to those for bottomonium
states.

\section{One-Loop QCD Calculation}

Now we proceed to calculate the one-loop QCD corrections. The
numerical calculation of one-loop QCD corrections is performed with
the help of \textbf{Feyncalc} and \textbf{Looptools}. At the one-loop
level of QCD, there are eight Feynman diagrams. We show four of them
in Fig.2, and the other four can be obtained by reversing the
direction of the charm quark line.
\begin{figure}[!]
\begin{center}
\scalebox{0.8}{\includegraphics{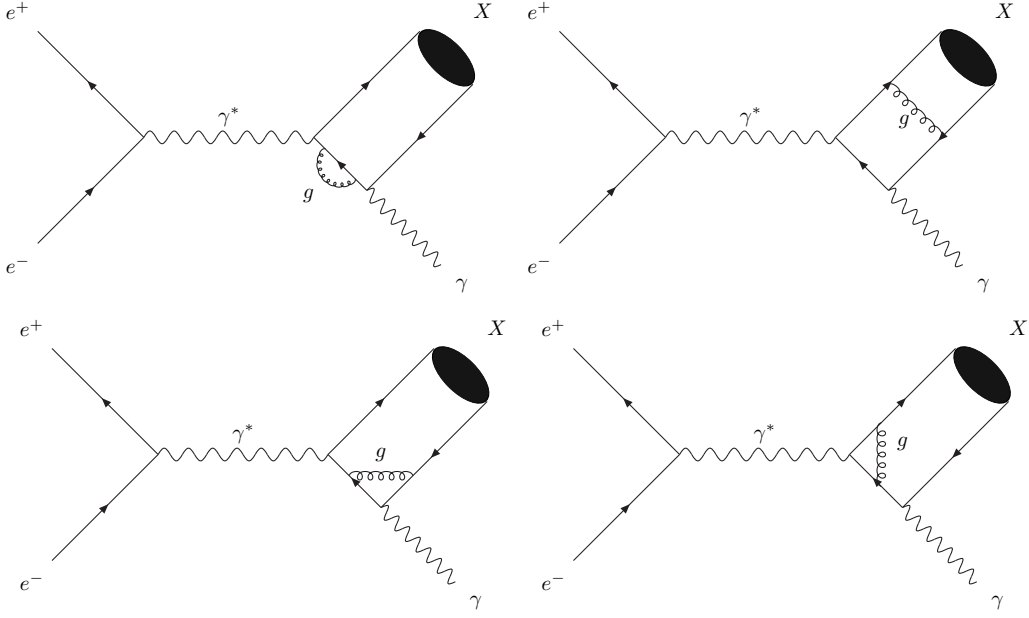}}
\end{center}
\vspace{-0.5cm} \caption{The one-loop QCD diagrams for
$e^+e^-\longrightarrow \gamma+X$}
\end{figure}
At order $\alpha^3\alpha_s$, the cross section for
$e^+e^-\longrightarrow \gamma+X$ is
\begin{eqnarray}
{\rm d}\sigma& \propto &|\mathcal{M}_{tree} + \mathcal{M}_{QCD}|^2 \nonumber \\
&=&|\mathcal{M}_{tree}|^2  \hspace{-0.1 cm}+ \hspace{-0.1 cm} 2 {\rm
Re}(\mathcal{M}_{tree}^*\mathcal{M}_{QCD}) \hspace{-0.1 cm}+
\hspace{-0.1 cm} {\cal{O}}(\alpha^3\alpha_s^2),
\end{eqnarray}
where $\mathcal{M}_{QCD}$ means the one-loop QCD amplitude. The
on-shell scheme is adopted and then the self-energy renormalization
constant $Z_1$ and vertex renormalization constant $Z_2$ are chosen
to be
\begin{eqnarray}
\delta Z_2^{\rm
OS}&=&-\frac{1}{\varepsilon_{UV}}+\gamma_E-4-\frac{2}{\varepsilon_{IR}}-\log(\frac{4\pi\mu^2}{m^2}),
\\\delta Z_1^{\rm OS}&=&\delta Z_2^{\rm OS},
\end{eqnarray}
where we omit the coefficient before the self-energy renormalization
constant and part of the infrared divergence term in $\delta
Z_2^{OS}$.

In the S-wave case, we encounter the $C_0$ function in box diagram,
with the analytic formula\cite{Zhang:2005cha}:
\begin{eqnarray}
&& C_0[p_{Q},-p_{\bar{Q}},0,m_Q,m_Q]=\nonumber
\\&&\frac{-i}{2m_Q^2(4\pi)^2}\left(\frac{4
\pi \mu^2}{m_Q^2}\right)^{\epsilon}\Gamma(1+\epsilon)\left[\,
\frac{1}{\epsilon} + \frac{\pi^2}{v} -2 \right].
\end{eqnarray}
The infrared divergence $\epsilon$ is canceled by the IR divergence
term in self-energy and vertex renormalization constants, and the
Coulomb singularity term with $\frac{1}{v}$ pole can be absorbed
into the wave-function by
\begin{eqnarray}
&&|R_s(0)|^2(1+A\frac{\alpha_s}{v}+B\alpha_s)\nonumber
\\&&=|R_s(0)|^2(1+A\frac{\alpha_s}{v})(1+B\alpha_s)+\mathcal{O}(\alpha_s^2).
\end{eqnarray}

In the P-wave case, we have to deal with loop-integrals typically as
the following expression in the box diagram when taking derivative
of the relative momentum $q_\alpha$ on the denominator of the
propagators
\begin{eqnarray}\int \hspace{-0.1cm}d^4
l\frac{A(l)l^\alpha}{l^2((\hspace{-0.05cm}l\hspace{-0.1cm}
-\hspace{-0.1cm}p_1\hspace{-0.05cm})^2\hspace{-0.1cm}-
\hspace{-0.1cm}m^2\hspace{-0.05cm})^2((l\hspace{-0.1cm}+
\hspace{-0.1cm}p_1)^2\hspace{-0.1cm}-\hspace{-0.1cm}m^2)
(\hspace{-0.05cm}(l\hspace{-0.1cm}-\hspace{-0.1cm}p_1\hspace{-0.1cm}-\hspace{-0.1cm}p_2)^2
\hspace{-0.1cm}-\hspace{-0.1cm}m^2\hspace{-0.05cm})}
\end{eqnarray}
where $2p_1$ is the momentum of the heavy quarkonium  and $p_2$ is
the momentum of the photon. The contribution which is proportional
to the $p_1^\alpha$ term will be omitted when contracted with
polarization vector. Using the identity $A(l)=(A(l)-A(0))+A(0)$, we
can separate the IR divergence into the second term which will be
canceled by other diagrams. Three types of integrations will appear
here, which are given in the \textbf{Appendix}.

In treating the first term, with the help of the formula $l\cdot
p_1=(l^2-((l-p_1)^2-m_Q^2))/2$ and Dirac decomposition: $l^\mu
l^\nu=g^{\mu\nu}I_1+p_1^\mu p_1^\nu I_2+(p_1^\mu p_2^\nu+p_1^\nu
p_2^\mu)I_3+p_2^\mu p_2^\nu I_4$, we are able to evaluate most of
terms by using \textbf{LoopTools}.  Note that to get the correct
result of the integration $\int d^4
l\frac{1}{((l-p_1)^2-m^2)^2((l+p_1)^2-m^2)}$ the small number
$i\varepsilon$ in the propagators should be kept. And we have
checked the independence of the final result on $\varepsilon$.

The analytical method is also performed to calculate the one-loop
QCD corrections as a cross-check for the numerical results. We find
the results of the two different methods are in agreement.

\section{Numerical result and Discussion}
We choose $\sqrt{s}=10.6$ GeV, $m_c$=1.5 GeV, $m_b$=4.7 GeV,
$\alpha_s(2m_c)$=0.26, $\alpha_s(2m_b)$=0.18 as inputs. As for the
charmonium wave-functions at the origin, we choose the results from
potential model calculations (see the results of the $B-T$-type potential
in Ref.\cite{23}), which are listed in Table I. The results of
cross sections for $e^+e^- \to\gamma^* \to\gamma+X$ are listed in
Table II, where $\sigma_{QED}$ means the QED result and $\sigma_{QCD}$
means the corresponding one-loop QCD correction. However, if we
extract the wave-functions at the origin from the observed
charmonium decay (e.g., $J/\psi\to e^+e^-$ or $\eta_c\to 2\gamma$)
widths using theoretical expressions with (without) NLO QCD
corrections\cite{20}, then the obtained QED cross sections for
$e^+e^{-}\to\gamma^* \to\gamma+X$ will be larger (smaller) than the
S-wave results given in Table II. These are the uncertainties due to
long-distance matrix elements, and our result in Table II is a rather
moderate one.\footnote{In Ref.\cite{Lee08} the authors get larger
values by using the $\eta_c\to 2\gamma$ width with NLO QCD
corrections as inputs.}
\begin{table}
\begin{center}
\caption{Numerical values of the radial wave functions at the origin
${|R_{nl}^{(l)}(0)|}^2$ for $c\bar{c}$ and $b\bar{b}$ calculated
with the QCD (BT) potential in Ref.\cite{23}.}
\begin{tabular}{|c|c|c|}
\hline States &\;$c\bar{c}$\;& \;$b\bar{b}$\;\\
\hline 1S & $0.81GeV^{3}$ & $6.477GeV^{3}$\\
\hline 2S & $0.529GeV^{3}$& $3.234GeV^{3}$\\
\hline 3S & $0.455GeV^{3}$&$2.474GeV^{3}$\\
\hline 1P & 0.075$GeV^{5}$& 1.417$GeV^{5}$\\
\hline 2P & 0.102$GeV^{5}$&--\\
\hline 1D & 0.015$GeV^{7}$&--\\
\hline
\end{tabular}
\label{table1}
\end{center}
\vspace{-0.5cm}
\end{table}

\begin{table}
\begin {center}
\caption{QED results for $e^+e^-\longrightarrow \gamma+X$ and the
one-loop QCD corrections with $m_c=1.5GeV$, $m_b=4.7GeV$,
$\alpha_s(2m_c)=0.26$, $\alpha_s(2m_b)=0.18$, where $\sigma_{QED}$
means the QED result and $\sigma_{QCD}$ means the corresponding
one-loop QCD correction. }
\begin{tabular}{|c|c|c|c|c|c|c|c|}
\hline
process&$\eta_c$&$\eta^{'}_c$&$\eta^{''}_c$&${}^1D_2$&$\eta_b$&$\eta^{'}_b$&$\eta^{''}_b$\\\hline
$\sigma_{QED}$(fb)&$59.1$&$38.6$&$33.2$&$1.08$&$2.19$&$0.16$&$0.01$\\
\hline $\sigma_{QCD}$(fb)&$-12.5$&$-8.19$&$-7.04$&$ $&$-0.55$&$\sim 0$&$\sim 0$\\
\hline
\end{tabular}
\begin{tabular}{|c|c|c|c|c|c|c|c|c|c|}
\hline
process&$\chi_{c0}$&$\chi_{c1}$&$\chi_{c2}$&$\chi^{'}_{c0}$&$\chi^{'}_{c1}$&$\chi^{'}_{c2}$
&$\chi_{b0}$&$\chi_{b1}$&$\chi_{b2}$\\\hline
$\sigma_{QED}$(fb)&$1.66$&$18.6$&$7.35$&$2.25$&$25.3$&$10.0$&$0.46$&$2.69$&$3.55$\\\hline
$\sigma_{QCD}$(fb)&$0.28$&$-5.13$&$-5.49$&$0.38$&$-6.98$&$-7.47$&$-0.20$&$-0.97$&$-1.38$\\\hline
\end{tabular}
\label{table2}
\end {center}
\vspace{-0.5cm}
\end{table}

We see that in most cases the one-loop QCD corrections are negative
and moderate, except for the $\chi_{c2}$ case, in which the
correction is large and is about  $-75\%$ of the QED result. This is
very different from the case of double charmonium production
$e^+e^-\to J/\psi~+~X$, where one-loop QCD corrections are positive
and large in most cases (see Refs.\cite{Braaten:2002fi} for LO and
Refs.\cite{Zhang:2005cha,Gong:2007db} for NLO corrections).

We find that one-loop QCD corrections do not change the angular
distributions of $\chi_{c0}$ and $\eta_c$, which read
$(1+\cos^2(\theta))$, confirmed by the effective Lagrangian method.
However, when including one-loop QCD corrections, the angular
distributions of $\chi_{c1}$ and $\chi_{c2}$ are changed from
$(1.38+\cos^2(\theta))$ and $(2.72+\cos^2(\theta))$ to
$(1.41+\cos^2(\theta))$ and $(1.61+\cos^2(\theta))$, which are shown
in Fig.[3] and Fig.[4], respectively.
\begin{figure}[!]
\begin{center}
\scalebox{0.6}{\includegraphics{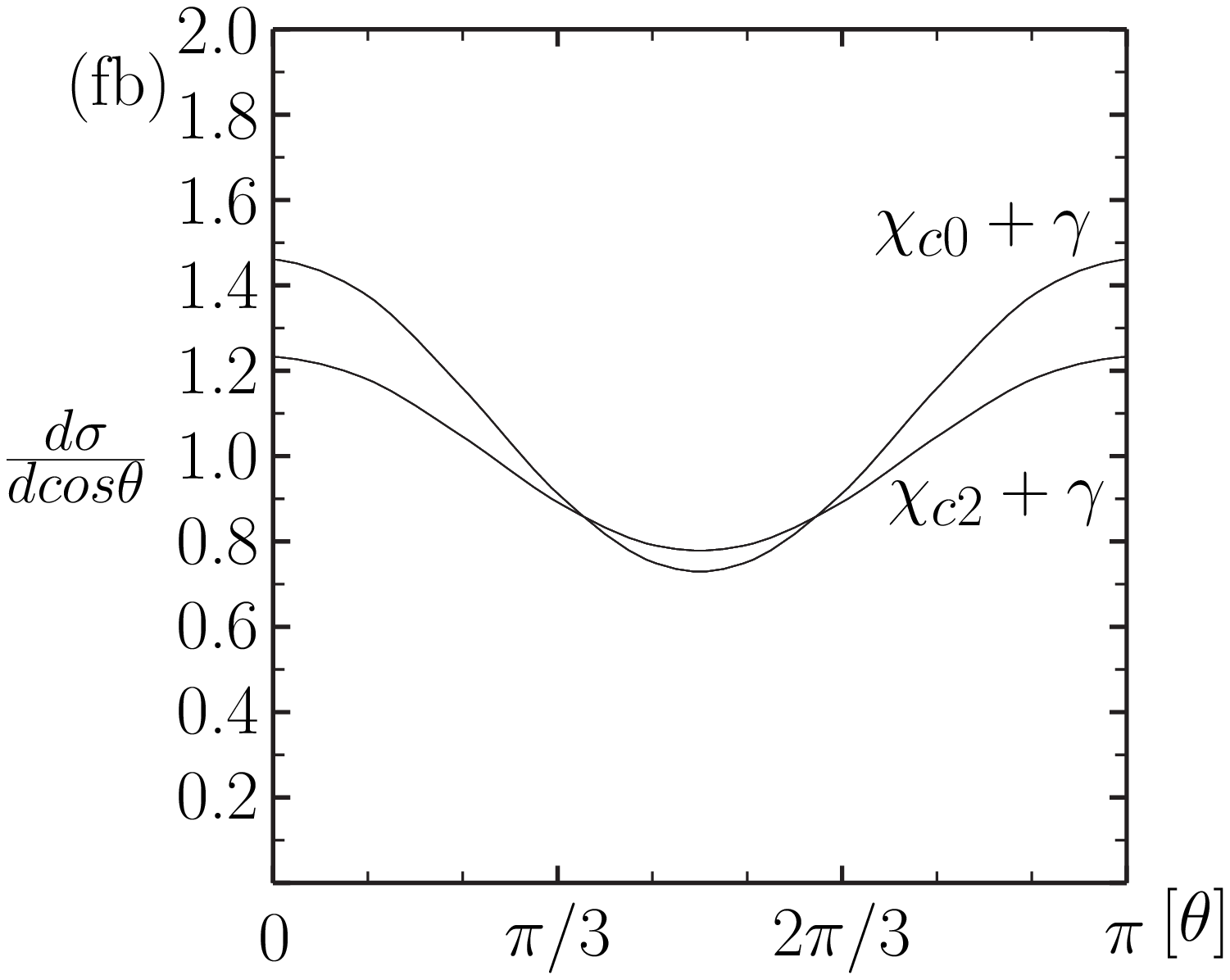}}
\end{center}
\caption{Angular distributions for $\chi_{c0}+\gamma$ and
$\chi_{c2}+\gamma$ productions in $e^{+}e^{-}$ annihilation up to
 order $\alpha^3\alpha_s$.}
\begin{center}
\scalebox{0.6}{\includegraphics{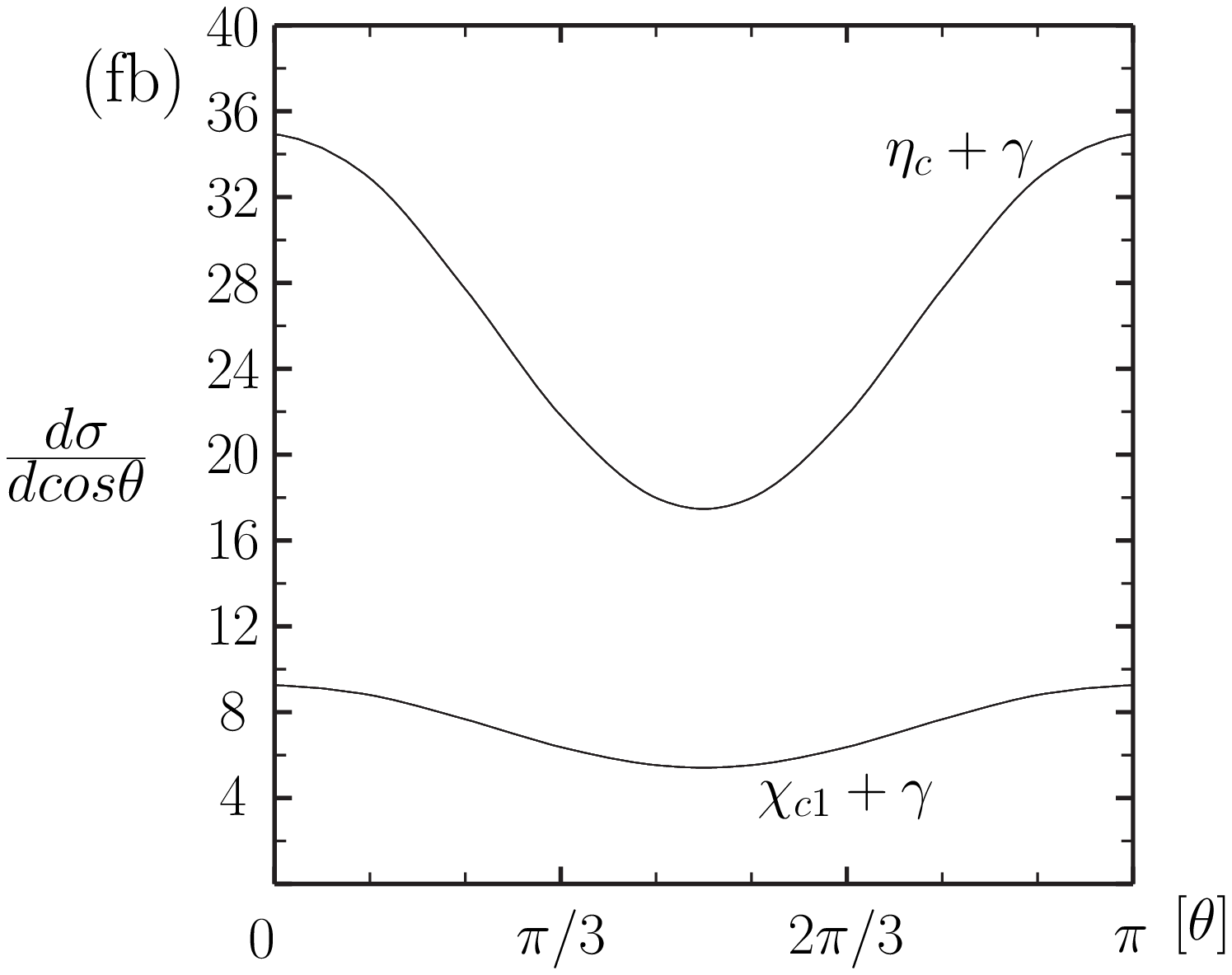}}
\end{center}
\caption{Angular distributions for $\chi_{c1}+\gamma$ and
$\eta_c+\gamma$ productions in $e^{+}e^{-}$ annihilation up to
 order $\alpha^3\alpha_s$.}
\end{figure}

\begin{table}
\begin{center}
\caption{Cross sections of QED with one-loop QCD corrections for
varying $\alpha_s$ and $m_c$.}
\begin{tabular}{|c|c|c|c|c|c|c|c|c|}
\hline - & \multicolumn{3}{|c|}{$\alpha_s=0.26$}&
\multicolumn{3}{|c|}{$\alpha_s=0.21$}\\
\hline $m_c (GeV)$ & $m_c=1.4$ & $m_c=1.5$ & $m_c=1.6$ & $m_c=1.4$ & $m_c=1.5$ & $m_c=1.6$ \\
\hline $\sigma(\eta_c)(fb)$ & 51.3 & 46.6 & 42.5 & 53.7
& 48.9 & 44.9 \\
\hline $\sigma(\eta_c^{'})(fb)$ & 33.5 & 29.4 & 27.7 & 35.1
& 31.4 & 29.3 \\
\hline $\sigma(\eta_c^{''})(fb)$ & 28.8 & 26.1 & 23.9 & 30.2 & 27.3
& 25.2
\\ \hline $\sigma(\chi_{c0})(fb)$ & 2.55 & 1.94 & 1.53 & 2.48
& 1.81 & 1.48 \\
\hline $\sigma(\chi_{c1})(fb)$ & 16.6 & 13.5 & 11.1 & 17.7 & 14.6 &
12.0 \\ \hline $\sigma(\chi_{c2})(fb)$ & 2.31 & 1.80 & 1.65& 3.53
& 2.86 & 2.55 \\
\hline $\sigma(\chi_{c0}^{'})(fb)$ & 3.46 & 2.63 & 2.07 & 3.36
& 2.56 & 2.01 \\
\hline $\sigma(\chi_{c1}^{'})(fb)$ & 22.6 & 18.4 & 15.1 & 24.1
& 19.7 & 16.3 \\
\hline $\sigma(\chi_{c2}^{''})(fb)$ & 3.14 & 2.53 & 2.25 & 4.80
& 3.97 & 3.48 \\
\hline
\end{tabular}

 \label{table3}
 \vspace{-6mm}
\end{center}
\end{table}

Since the production of $b\overline{b}$ mesons is near threshold, we
make up a factor in the phase space by
$\frac{(1-(M^2_{\eta_b}/s)^2)^3}{(1-(4m^2_b/s)^2)^3}$ for $\eta_b$
production (P-wave process), and by
$\frac{1-(M^2_{\chi_{bJ}}/s)^2}{1-(4m^2_b/s)^2}$ for $\chi_{bJ}$
production (S-wave dominated process) as a rough remedy to the phase
space integrals. The states of $\eta_b^\prime$ and
$\eta_b^{\prime\prime}$ have not been observed yet, so we use the
masses of observed $\Upsilon(2S)$ and $\Upsilon(3S)$ for
replacement. Because of the suppression from the small phase space, the
cross sections for $\eta_b^\prime$, $\eta_b^{\prime\prime}$ and
$\chi_{b0}$ are negligible and not useful phenomenologically. We
also choose different values of $\alpha_s$ and $m_c$ as inputs for
comparison, and the obtained cross sections of QED with one-loop QCD
corrections are shown in Table III.

From our results, we see the production cross sections for $\eta_c$,
$\eta_c^\prime$, and $\eta_c^{\prime\prime}$ are about
$47\mathrm{fb}$, $29\mathrm{fb}$, and $26\mathrm{fb}$ respectively
for $m_c=1.5\mathrm{GeV}$ and $\alpha_s=0.26$. Since the new state
X(3940), which is seen in the spectrum recoiling against the
$J/\psi$ in the inclusive process $e^+e^-\longrightarrow
J/\psi+anything$ by Belle\cite{Abe:2007jn}, is widely believed to be
the $\eta_c^{\prime\prime}$ state (see, e.g., \cite{Li09} for
discussions), we expect it could be seen in the recoil spectrum
against the photon. We also find that the cross sections for
$\chi_{c1}$ and $\chi_{c1}^\prime$ are 13fb and 18fb respectively,
which are much larger than that produced in the double charmonium
process $e^+e^-\longrightarrow J/\psi+\chi_{c1}(\chi_{c1}^\prime)$
recoiling against $J/\psi$. As long as the background of ISR can be
largely removed, the exclusive process $e^+e^-\longrightarrow
\gamma+\chi_{c1}(\chi_{c1}^\prime)$ can be an alternative probe to
the $\chi_{c1}$ meson as well as $\chi_{c1}(2P)$ meson. It will be
interesting to see whether there will be signals of X(3872) or
X(3940) as the candidates of $\chi_{c1}(2P)$. Whereas the predicted
production rates of $\chi_{c0}$ and its radial excitations in
$e^+e^-\longrightarrow \gamma+\chi_{c0}(\chi_{c0}^\prime)$ are much
smaller than that in $e^+e^-\to J/\psi+\chi_{c0}(\chi_{c0}^\prime)$.
By comparing the measurements of the $J/\psi$ process with the
photon process, we may clarify whether the X(4160), which is
copiously produced in association with $J/\psi$, is a radially
excited state of $\chi_{c0}$ (say $\chi_{c0}(3P)$), or it is the
radial excitation of $\eta_c$ (say $\eta_c(3S)$)(see Ref.\cite{Li09}
for discussions on the X(4160)).

We see that although the photon and $J/\psi$ meson have the same
quantum number of $J^{PC}=1^{--}$, when P-wave charmonium states are
produced in association with the photon or $J/\psi$ in $e^{+}e^{-}$
annihilation, the behaviors of $e^{+}e^{-}\to J/\psi+X$ and
$e^{+}e^{-}\to\gamma+X$ are very different. In the $J/\psi$ case,
the associated production of $\chi_{c0}$ state is prominent, whereas
the photon favors  being associated with the $\chi_{c1}$ state.
Hopefully, the measurement of structures recoiling against the
photon in the $e^+e^-\longrightarrow \gamma+ X$ process, especially
via the exclusive channels $J/\psi\pi^+\pi^-, \psi(2S)\gamma,
J/\psi\gamma$ and $J/\psi\omega$, will provide a possible way to
search for the new heavy quarkonium states,  when more experimental
data are accumulated in the future, and the background from ISR
process is largely removed.

\section*{\large{Acknowledgments}}

We thank Xin-Chou Lou, Chang-Zheng Yuan, and Chang-Chun Zhang for
useful discussions concerning the experimental measurements at B
factories, and Ce Meng and Yu-Jie Zhang for helpful discussions. One
of us (D.L.) would like to thank Jun Se for a useful suggestion in a
numerical calculation.  This work was supported by the National
Natural Science Foundation of China (No 10675003, No 10721063) and
the Ministry of Science and Technology of China (2009CB825200).
Zhi-Guo He is currently supported by the Ministry of Science and
Innovation of Spain (Contract No.CPAN08-PD14 of the CSD2007-00042
Consolider-Ingenio 2010 program, and the FPA2007-66665-C02-01/
project (Spain)).

$Note$. After this work was completed, we learned a similar work was
done by Sang and Chen\cite{SC}, and their result is consistent with
ours.

\section*{\large{Appendix}}

When we evaluate the numerical result, there are some basic loop
integrals. They are given by
\begin{subequations}
\begin{eqnarray}
Im\int d^4 l\frac{(l\cdot
p_2)^2}{l^2((l-p_1)^2-m_c^2)^2((l+p_1)^2-m_c^2)((l-p_1-p_2)^2-m_c^2)}=4
\end{eqnarray}
\begin{eqnarray}
Im\int d^4 l\frac{(l\cdot
p_2)^2}{l^2((l-p_1)^2-m_c^2)((l+p_1)^2-m_c^2)^2((l-p_1-p_2)^2-m_c^2)}=-14.7
\end{eqnarray}
\begin{eqnarray}
Im\int d^4 l\frac{l\cdot
p_2}{l^2((l-p_1)^2-m_c^2)((l+p_1)^2-m_c^2)((l-p_1-p_2)^2-m_c^2)^2}=-0.0786
\end{eqnarray}
\end{subequations}
where we chose $s=10.6^2GeV^2, m_c=1.5GeV$.



\begin{thebibliography}{}






\bibitem{rev} S.L.~Olsen, arXiv:0801.1153;~ S. Godfrey and S.L. Olsen,
arXiv:0801.3867; ~E.S.~Swanson, Phys. Rept. 429, 243 (2006).

\bibitem{10}
E.S. Swanson, Phys. Lett. \textbf{B588}, 189 (2004); ibid
\textbf{B598}, 197 (2004).


\bibitem{11}
X. Liu, B. Zhang and S.L. Zhu, Phys. Lett. \textbf{B644} 355,
(2007).


\bibitem{12}
C. Meng and K.T. Chao. Phys. Rev. \textbf{D75}, 114002 (2007); C.
Meng, Y.J. Gao, and K.T. Chao, arXiv:hep-ph/0502240.

\bibitem {Li09}B.~Q.~Li and K.~T.~Chao, Phys. Rev. D79, 094004 (2009); K.T.
Chao, Phys. Lett. B661, 348 (2008).



\bibitem{Abe:2002rb}
K.~Abe {\it et al.} [BELLE Collaboration],
Phys.\ Rev.\ Lett.\ {\bf 89}, 142001 (2002);
%
%
%
K.~Abe {\it et al.}[Belle Collaboration],
Phys.Rev. D70 (2004) 071102.

\bibitem{BaBar:2005}
  B.~Aubert {\it et al.}  [BABAR Collaboration],
  Phys.\ Rev.\ D {\bf 72}, 031101 (2005).

\bibitem{Lee08}
H.S. Chung, J. Lee and C. Yu, Phys. Rev. D78, 074022 (2008)
[arXiv:0808.1625].


\bibitem{Bodwin:1994jh}
G.T.~Bodwin, E.~Braaten, and G.P.~Lepage,
    Phys.\ Rev.\ D {\bf 51}, 1125 (1995);
{\bf 55}, 5853(E) (1997).



\bibitem{Braaten:2002fi}
  E.~Braaten and J.~Lee,
  Phys.\ Rev.\  D {\bf 67}, 054007 (2003)
  [Erratum-ibid.\  D {\bf 72}, 099901 (2005)];
  K.~Y.~Liu, Z.~G.~He and K.~T.~Chao,
  Phys.\ Lett.\  B {\bf 557}, 45 (2003),
  Phys.\ Rev.\  D {\bf 77}, 014002 (2008);
K. Hagiwara, E. Kou, and C.F. Qiao, Phys. Lett. B570, 39 (2003).


\bibitem{Zhang:2005cha}
  Y.~J.~Zhang, Y.~J.~Gao and K.~T.~Chao,
  Phys.\ Rev.\ Lett.\  {\bf 96}, 092001 (2006).


\bibitem{Gong:2007db}
 B.~Gong and J.~X.~Wang,
  Phys.\ Rev.\  D {\bf 77}, 054028 (2008).


\bibitem{bodwin06}
G.T. Bodwin, D. Kang and J. Lee, Phys. Rev.\textbf{D74}, 014014
(2006); Phys. Rev. D74, 114028 (2006); G.T. Bodwin, J. Lee, and C.
Yu, Phys. Rev. D77, 094018 (2008).


\bibitem{He:2007te}
  Z.~G.~He, Y.~Fan and K.~T.~Chao,
  Phys.\ Rev.\  D {\bf 75}, 074011 (2007).

\bibitem{Shifman:1980dk}
  M.~A.~Shifman and M.~I.~Vysotsky,
  Nucl.\ Phys.\  B {\bf 186}, 475 (1981).



\bibitem{jia} Y. Jia and D.S. Yang, Nucl. Phys. B814, 217 (2009).


\bibitem{Kuhn:1979bb}
  J.~H.~Kuhn, J.~Kaplan and E.~G.~O.~Safiani,
  Nucl.\ Phys.\  B {\bf 157}, 125 (1979);
  B.~Guberina, J.~H.~Kuhn, R.~D.~Peccei and R.~Ruckl,
  Nucl.\ Phys.\  B {\bf 174}, 317 (1980);
  E.~L.~Berger and D.~L.~Jones,
  Phys.\ Rev.\  D {\bf 23}, 1521 (1981).

\bibitem{20}
G.T. Bodwin and A. Petrelli, Phys. Rev. \textbf{D66}, 094011(2002).


\bibitem{Ko:1996xw}
  P.~Ko, J.~Lee and H.~S.~Song,
  Phys.\ Rev.\  D {\bf 54}, 4312 (1996)
  [Erratum-ibid.\  D {\bf 60}, 119902 (1999)]
  [arXiv:hep-ph/9602223].


\bibitem{Braaten:1995ez}
  E.~Braaten and Y.~Q.~Chen,
  Phys.\ Rev.\ Lett.\  {\bf 76}, 730 (1996)
  [arXiv:hep-ph/9508373].

\bibitem{23}
E.J. Eichten and C. Quigg, Phys. Rev. \textbf{D52}, 1726 (1995).

\bibitem{Abe:2007jn}
  K.~Abe {\it et al.},
  Phys.\ Rev.\ Lett.\  {\bf 98}, 082001 (2007)
  [arXiv:hep-ex/0507019];
  P.~Pakhlov {\it et al.}  [Belle Collaboration],
  Phys.\ Rev.\ Lett.\  {\bf 100}, 202001 (2008)
  [arXiv:0708.3812 [hep-ex]].


\bibitem{SC} W.L. Sang and Y.Q. Chen, arXiv:0910.4071.










































\end{thebibliography}
\end{document}